\documentclass[twocolumn,showpacs]{revtex4-1}

\usepackage{newlfont}
\usepackage[below]{placeins}
\usepackage{flafter}
\usepackage{amsfonts}
\usepackage{amsmath}
\usepackage{amssymb}
\usepackage[american]{babel}
\usepackage{graphicx}
\usepackage{multirow}
\usepackage{url}
\usepackage[linesnumbered,ruled,vlined]{algorithm2e}
\usepackage{color}
\usepackage{mathtools}

\allowdisplaybreaks

\newcommand{\ket}[1]{\big| #1 \big\rangle}
\newcommand{\bra}[1]{\big\langle #1 \big|}
\newcommand{\braket}[2]{\big\langle #1 \big| #2 \big\rangle}                 
\newcommand{\bracket}[3]{\big\langle #1 \big| #2 \big| #3 \big\rangle}       

\newtheorem{theorem}{Theorem}[section]
\newtheorem{lemma}[theorem]{Lemma}

\newcommand{\qed}{\nobreak \ifvmode \relax \else
      \ifdim\lastskip<1.5em \hskip-\lastskip
      \hskip1.5em plus0em minus0.5em \fi \nobreak
      \vrule height0.75em width0.5em depth0.25em\fi}

\begin{document}


\title{Staggered Quantum Walk on Hexagonal Lattices}
\author{ Bruno Chagas$^1$, Renato Portugal$^1$, Stefan Boettcher$^2$, Etsuo Segawa$^3$}
\affiliation{ 
$^1$National Laboratory of Scientific Computing (LNCC), Petr\'{o}polis, RJ,  25651-075, Brazil\\
$^2$Phys Dept, Emory University, USA\\
$^3$Tohoku University, Japan
}

\date{\today}

\begin{abstract}
A discrete-time staggered quantum walk was recently introduced as a generalization that allows to unify other versions, such as the coined and Szegedy's walk. However, it also produces new forms of quantum walks not covered by previous versions. To explore their properties, we study here the staggered walk on a hexagonal lattice. Such a walk is defined using a set of overlapping tessellations that cover the graph edges, and each tessellation is a partition of the node set into cliques. The hexagonal lattice requires at least three tessellations. Each tessellation is associated with a local unitary operator and the product of the local operators defines the evolution operator of the staggered walk on the graph. After defining the evolution operator on the hexagonal lattice, we analyze the quantum walk dynamics with the focus on the position standard deviation and localization. We also obtain analytic results for the time complexity of spatial search algorithms with one marked node using cyclic boundary conditions.
\end{abstract}


\maketitle

\section{Introduction}

The discrete-time coined quantum walk on graphs, which is the quantum version of the classical random walk, was introduced by Aharonov \textit{et al.}~\cite{AAKV01}. In this model, the vertices are the possible locations of the walker, who is endowed with an internal state that determines the direction of the motion. The evolution operator is the product of a shift operator, which moves the walker to an adjacent vertex, and the coin operator, which acts in the internal space. Coined walks admit the phenomenon of localization~\cite{IKS05,Che16,VFQF17,Der18}. There are many experimental proposals of the coined model~\cite{MW13} and actual realizations in laboratories~\cite{BLZTX17}.


Quantum walks on two-dimensional lattices are interesting for many a reason. The dynamics is richer than the one on the line, yet amenable to analytic calculations. Quantum spatial search on finite two-dimensional lattices is the physically most convenient geometry to realize Grover's quantum search algorithm~\cite{Gro97a} with nearly optimal quantum gain~\cite{AKR05}. There are three kinds of two-dimensional regular lattices: square, hexagonal, and triangular.
All these lattices have attracted the attention of researchers. 
For instance, coined quantum walks were analyzed on the two-dimensional square lattice~\cite{AAA10,XZBZQS15,CA16,MGR16,KK17,EKOS17}, hexagonal lattice~\cite{VAL13,LLS15,Bohetal15,AMMP18}, and triangular lattice~\cite{ADFP12,MOS16,AMMP18}. Since the seminal paper by Shenvi \textit{et al.}~\cite{SKW03}, coined quantum walks are being used for search algorithms~\cite{Por13}. Search on the two-dimensional square lattice was analyzed in~\cite{AKR05,Tul08,won17}. Search on the hexagonal lattice was analyzed in~\cite{ADFP12}, which showed that the optimal number of steps is $O(\sqrt{N\ln N})$ and the success probability is $O(1/\ln N)$, where $N$ is the number of sites, and quantum transport on the hexagonal lattice was analyzed in~\cite{BACK16}. Search on the triangular lattice was analyzed in~\cite{ADFP12,MOS17}.

\begin{figure}[h!] 
\centering
\includegraphics[trim=120 80 20 5,clip,scale=0.11]{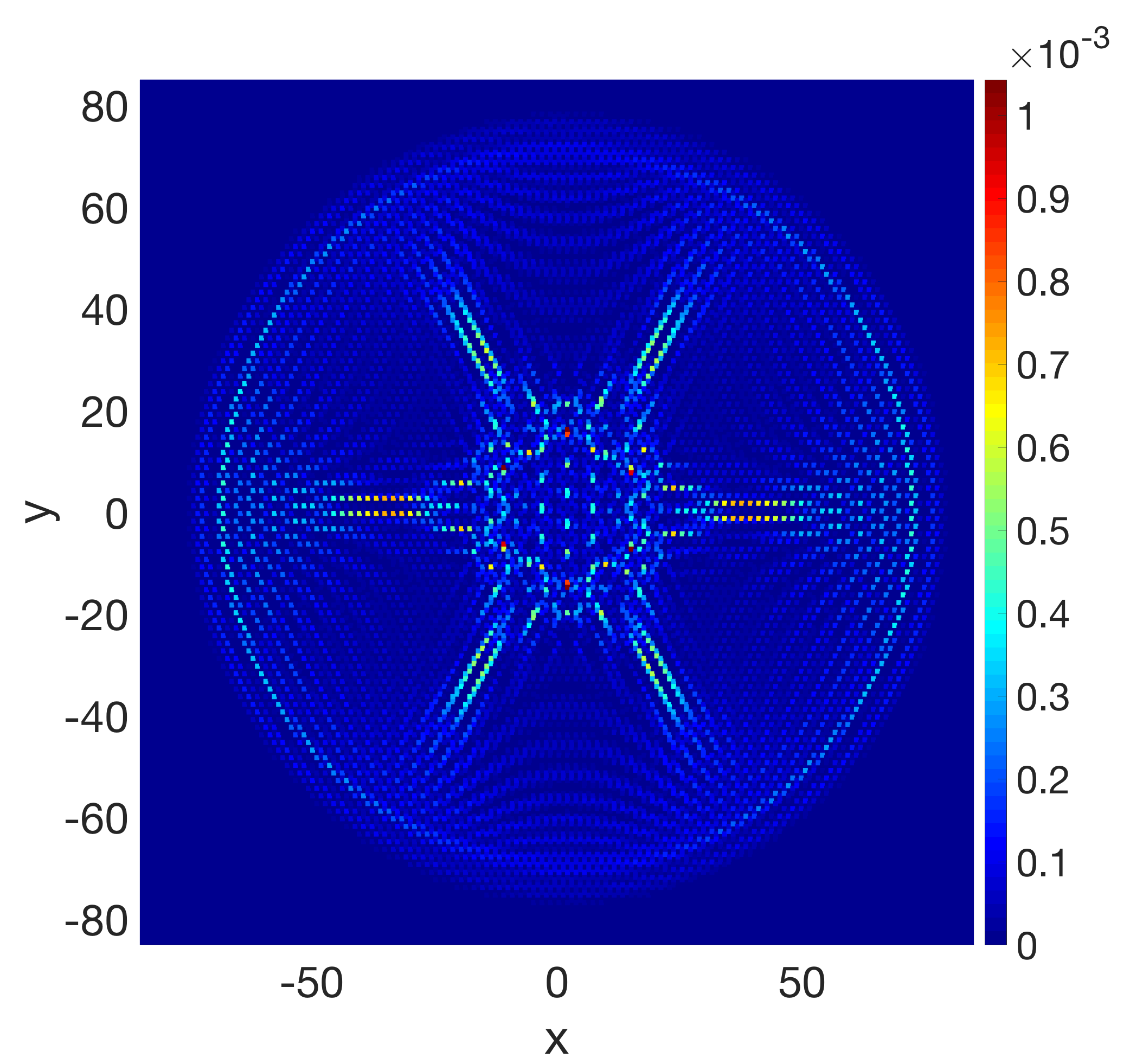}
\caption{Probability distribution of a staggered quantum walk on the hexagonal lattice after 58 steps.}
\label{fig:prob_dist_intro}
\end{figure}


Patel \textit{et al.}~\cite{PRR05} showed that the coin is not necessary for a quantum walk on the line and Portugal \textit{et al.}~\cite{PSFG16} described a complete model, called staggered quantum walk, on arbitrary graphs without using coins. In the staggered model, the evolution operator is a product of local unitary operators, which are obtained from graph tessellations. A tessellation is a partition of the vertex set into polygons, which are sets of adjacent vertices. Each polygon is associated with a unit vector in the Hilbert space spanned by the vertices of the polygon and a tessellation is associated with a local unitary operator. An experimental proposal of staggered quantum walks on triangle-free graph class, which includes hypercubic lattices, was proposed in~\cite{MOP17}. Quantum search on two-dimensional square lattices was addressed in~\cite{PF17} and the boundary-induced coherence on different topologies was addressed in~\cite{MR18}.

In this work, we focus our attention on the staggered quantum walk with Hamiltonians on the hexagonal lattice (or honeycomb network). Fig.~\ref{fig:prob_dist_intro} depicts an example of the probability distribution of a staggered quantum walk on the hexagonal lattice after 58 steps using an initial state with uniform amplitudes at two neighboring nodes. Note that the signature of the walk is different from the coined quantum walk.  Besides, there is no localization when we use the staggered walk on the hexagonal lattice, different from the coined walk, which makes localization on any 2D lattice almost inevitable~\cite{HKSS14,LYW15}.

 The first step is to define the evolution operator, which is based on a tessellation cover~\cite{POM17}.  Because the hexagonal lattice is a triangle-free graph, the tessellation cover can be obtained from a proper edge-coloring (see Fig.~\ref{fig:honeycomb1}). Each tessellation contains edges of the same color and comprises 2-vertex sets, whose union is the entire vertex set. Each tessellation is associated with a Hamiltonian $H_j$ $(0\le j\le 2)$ and with a local unitary operator $\exp(i\theta H_j)$, where $\theta$ is an angle. The evolution operator is the product of these unitary operators. After defining the evolution operator, we analyze numerically the mean square displacement, which depends on $\theta$. Starting at the origin, the fastest spread is obtained with $\theta=\pi/3$. 
 
We also address the spatial search problem on a $N$-vertex hexagonal lattice with cyclic boundary conditions. We prove analytically that a marked vertex can be found after $O(\sqrt{N\ln N})$ steps with success probability $O(1/\ln N)$, which can be improved to  $O(\sqrt{N\ln N})$ steps with success probability $O(1)$ after using Tulsi's modification~\cite{Tul12}. This result is obtained only if $\theta=\pi/3$, which coincides with the quickest spreading rate.

The structure of this paper is as follows. Sec.~\ref{sec:evol} describes the evolution operator of the staggered quantum walk on the hexagonal lattice. Sec.~\ref{sec:FB} obtains the spectral decomposition of the evolution operator using the staggered Fourier transform. Sec.~\ref{sec:SD} analyzes numerically the position standard deviation of the quantum walk. Sec.~\ref{sec:localization} shows that the staggered walk does not admit localization on the hexagonal lattice. Sec.~\ref{sec:search} presents the calculation of the time complexity of the spatial search algorithm with one marked node. Sec.\ref{sec:conc} describes our conclusions.

\section{The evolution operator for the hexagonal lattice}\label{sec:evol}

\begin{figure}[h!] 
\centering
\includegraphics[trim=265 0 0 0,clip,scale=0.4]{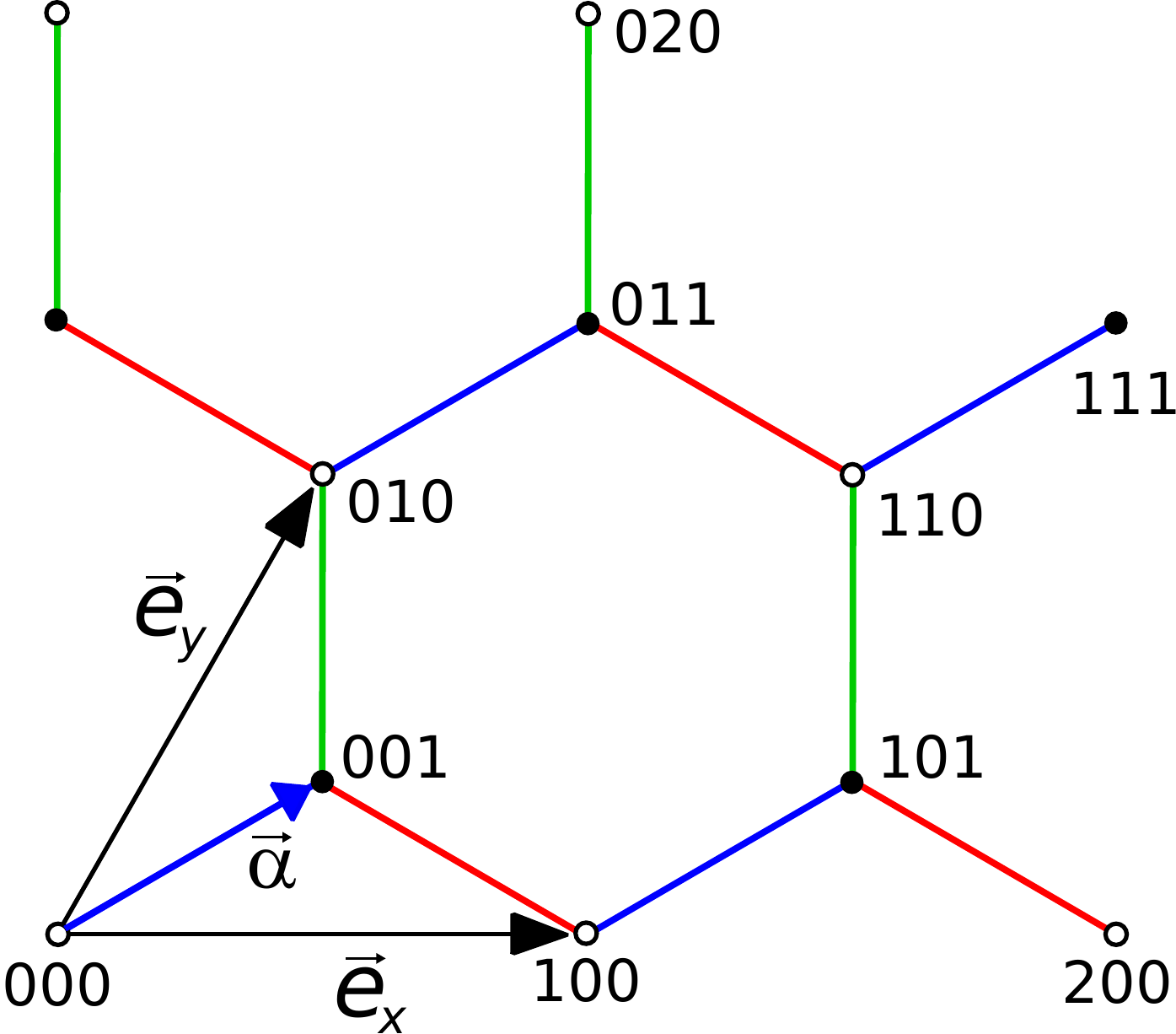}
\caption{The figure shows a representative part of a hexagonal lattice. Vectors $\vec e_x$ and $\vec e_y$ are shown in black and the vector $\vec \alpha$ in blue. The boundary conditions are cyclic following the directions $\vec e_x$ and $\vec e_y$. The vertices have labels $(x,y,i)$, where $i=0$ for the empty vertices and $i=1$ for the full vertices.  The  tessellation cover is depicted using colors red, green, and blue.}
\label{fig:honeycomb1}
\end{figure}

Fig.~\ref{fig:honeycomb1} depicts part of a hexagonal lattice. The vertex labels are chosen using a method similar to the one used for the two-dimensional square lattice. Recall that for the square lattice, the vertex label $(x,y)$ means that it is represented by vector $x\vec e_x+y\vec e_y$, where $\vec e_x$ and $\vec e_y$ are the orthogonal canonical vectors along axes $x$ and $y$, respectively. For the hexagonal lattice, the unit canonical vectors must be replaced by the non-orthogonal vectors $\vec e_x$ and $\vec e_y$ (in black) displayed in Fig.~\ref{fig:honeycomb1}. The vertices are split into two sets of equal cardinality: empty and full vertices. The positions of the empty vertices are obtained using vectors $x\vec e_x+y\vec e_y$, for $0\le x,y < n$, where $n$ is the even number of hexagons in the $x$- or $y$-directions (we are using the cyclic or torus-like boundary conditions). The total number of vertices is $N=2n^2$. The full vertices are obtained using vectors  $x\vec e_x+y\vec e_y+\vec \alpha$, where $\vec \alpha=(\vec e_x+\vec e_y)/3$, as shown in Fig.~\ref{fig:honeycomb1}. So, we use labels $(x,y,0)$ for the empty vertices and $(x,y,1)$ for the full vertices.

The tessellation cover is depicted using colors red, green, and blue in Fig.~\ref{fig:honeycomb1}. The tessellations are the following sets of 2-vertex sets
\begin{eqnarray*}
{\mathcal{T}}_\textrm{red}&=&\big\{ \{(x,y,1),(x+1,y,0)\}:\,0\le x,y<n\big\},\\
{\mathcal{T}}_\textrm{green}&=&\big\{ \{(x,y,1),(x,y+1,0)\}:\,0\le x,y<n\big\},\\
{\mathcal{T}}_\textrm{blue}&=&\big\{ \{(x,y,0),(x,y,1)\}:\,0\le x,y<n\big\},
\end{eqnarray*}
where the arithmetic is performed modulo $n$ (cyclic boundary conditions). Notice that each tessellation includes all vertices and the tessellation union covers all edges. The tessellations are associated with the following sets of vectors ($0\le x,y<n$):
\begin{eqnarray}\label{alpha_xyi}
\ket{\eta^0_{x,y}}&=&\frac{1}{\sqrt 2}\big(\ket{x,y,1}+\ket{x+1,y,0}\big),\nonumber\\ 
\ket{\eta^1_{x,y}}&=&\frac{1}{\sqrt 2}\big(\ket{x,y,1}+\ket{x,y+1,0}\big),\\ 
\ket{\eta^2_{x,y}}&=&\frac{1}{\sqrt 2}\big(\ket{x,y,0}+\ket{x,y,1}\big),\nonumber
\end{eqnarray}
and each set of vectors are used to define the Hermitian operators
\begin{eqnarray}\label{H0H1H2}
H_0&=&2\sum_{x,y=0}^{n-1}\ket{\eta^0_{x,y}}\bra{\eta^0_{x,y}}-I,\nonumber\\ 
H_1&=&2\sum_{x,y=0}^{n-1}\ket{\eta^1_{x,y}}\bra{\eta^1_{x,y}}-I,\\ 
H_2&=&2\sum_{x,y=0}^{n-1}\ket{\eta^2_{x,y}}\bra{\eta^2_{x,y}}-I,\nonumber
\end{eqnarray}
which act on the Hilbert space  ${\mathcal{H}}^{N}$. The evolution operator is
\begin{equation}\label{eq:U}
	U\,=\,\textrm{e}^{i\theta_2 H_2}\,\textrm{e}^{i\theta_1 H_1}\,\textrm{e}^{i\theta_0 H_0},
\end{equation}
where $\theta_0$, $\theta_1$, and $\theta_2$ are angles. In this work we consider $\theta_0=\theta_1=\theta_2=\theta$.

\begin{figure}[h!] 
\centering
\includegraphics[trim=80 70 20 5,clip,scale=0.2]{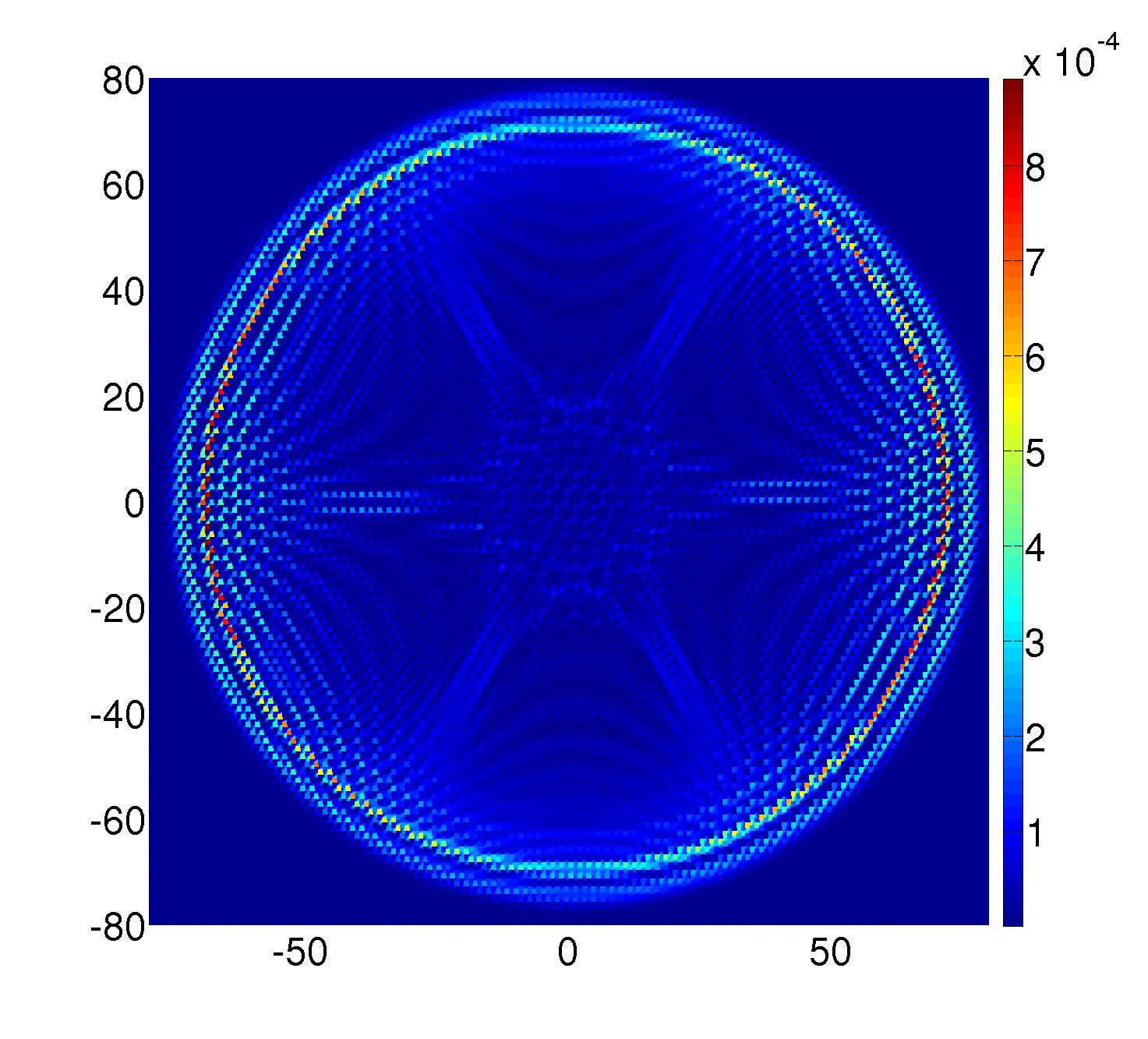}
\caption{Probability distribution after 58 steps with $\theta=\pi/3$ and using the initial state with uniform amplitudes in the central hexagon of Fig.~\ref{fig:honeycomb1}.}
\label{fig:prob_dist}
\end{figure}

If the initial condition is $\ket{\psi(0)}$, the state of the quantum walk after $t$ time steps is $\ket{\psi(t)}=U^t\ket{\psi(0)}$. The probability of finding the walker on node $\ket{x,y,i}$ after $t$ steps is $p_{x,y,i}(t)=|\braket{x,y,i}{\psi(t)}|^2$. When we fix $t$, $p_{x,y,i}(t)$ is a probability distribution. An example of the probability distribution after $t=58$ steps is depicted in Fig.~\ref{fig:prob_dist}, which used the initial condition
\begin{align}
 \ket{\psi(0)}=\frac{1}{\sqrt 6}\big(\ket{1,1,0}+\ket{1,0,1}+
 \ket{1,0,0}+ \nonumber\\
 \ket{0,0,1}+
 \ket{0,1,0}+\ket{0,1,1}\big),
\end{align} 
and $\theta=\pi/3$. Note that the highest values (red color) of the probability is close to the boundary of the physically achievable area, showing that the wave function spreads faster than the one that produces the distribution of Fig.~\ref{fig:prob_dist_intro}, whose initial condition is $\big(\ket{1,1,0}+\ket{1,0,1}\big)/\sqrt{2}$. 


\section{Fourier Basis}\label{sec:FB}

Notice that the hexagonal lattice is invariant under translations  $n_1\vec e_x+n_2\vec e_y+n_3\vec \alpha$, where $n_1,n_2,n_3$ are integers. Using the partition described in Sec.~\ref{sec:evol}, we add up the vertices in each class (empty and full vertices) using the Fourier amplitudes in the following way:
\begin{eqnarray}
 \ket{\psi_{k \ell}^0}&=&\frac{1}{n}\sum_{x,y=0}^{n-1}\omega^{kx+\ell y}\ket{x,y,0},\label{psikl0}\\
 \ket{\psi_{k \ell}^1}&=&\frac{1}{n}\sum_{x,y=0}^{n-1}\omega^{kx+\ell y}\ket{x,y,1},\label{psikl1}
\end{eqnarray}
where $\omega=\exp(2\pi i/n)$, $0\le k,\ell\le n-1$. Vectors $\ket{\psi_{k \ell}^0}$ and $\ket{\psi_{k \ell}^1}$ for all $k,l$ form an orthonormal Fourier basis of ${\mathcal{H}}^{N}$.

Now we show that, after fixing $k$ and $\ell$, the subspace spanned by $\ket{\psi_{k \ell}^0}$ and $\ket{\psi_{k \ell}^1}$ is invariant under the action of $U$. In fact, we obtain
\begin{eqnarray}
U\ket{{\psi}_{k \ell}^{\textrm{0}}} & = &A_{k \ell}\ket{{\psi}_{k \ell}^{\textrm{0}}}- B_{k \ell}^*\ket{{\psi}_{k \ell}^{\textrm{1}}},\label{eq:Hpsitilde1}\\
U\ket{{\psi}_{k \ell}^{\textrm{1}}} & = &B_{k \ell}\ket{{\psi}_{k \ell}^{\textrm{0}}}+A_{k \ell}^*\ket{{\psi}_{k \ell}^{\textrm{1}}},\label{eq:Hpsitilde2}
\end{eqnarray}
where 
\begin{eqnarray}
A_{k \ell}&=&(a_{k \ell}+i\,b_{k \ell})\cos{\theta},\\
B_{k \ell}&=&(c_{k \ell}+i\,d_{k \ell})\sin{\theta},
\end{eqnarray}
and\begin{eqnarray}
a_{k \ell}&=& - f_{k\ell}\sin^2{\theta}+\cos^2{\theta}    ,\label{a_kl}\\ 
b_{k \ell}&=&  - \,g_{k\ell}\sin^2{\theta},\label{b_kl}\\
c_{k \ell}&=& b_{k \ell}+\sin{\tilde{k}}+\sin{\tilde{\ell}},\\
d_{k \ell}&=& a_{k \ell}+\cos{\tilde{k}}+\cos{\tilde{\ell}},
\end{eqnarray}
where 
\begin{eqnarray}
f_{k \ell}&=&\cos{\tilde{k}}+\cos{\tilde{\ell}}+\cos{(\tilde{k}-\tilde{\ell})},\\
g_{k \ell}&=&\sin{\tilde{k}}+\sin{\tilde{\ell}}+\sin{(\tilde{k}-\tilde{\ell})}.
\end{eqnarray}
The new tilde variables  are $\tilde{k}={2\pi k}/{n}$ and $\tilde{\ell}={2\pi \ell}/{n}.$

The analysis of the dynamics is reduced to a two-dimensional space by defining a reduced evolution operator 
\begin{equation}
U_{k \ell}=\left[\begin{array}{cc}
 A_{k \ell} & B_{k \ell} \\
  - B_{k \ell}^* &  A_{k \ell}^*
\end{array}\right],
\end{equation}
which is unitary because $\left|A_{k \ell}\right|^2+\left|B_{k \ell}\right|^2=1.$ The eigenvalues of $U_{k \ell}$ are $\textrm{e}^{\pm i\phi_{k\ell}}$ where $\cos\phi_{k\ell}=a_{k\ell}\cos{\theta}.$ The eigenvectors are
\begin{equation}
 \ket{v_{k \ell}^{\pm\phi}}\,=\, \frac{1}{\sqrt{\gamma^\pm}}\left[ \begin {array}{c} B_{{k\ell}}\\ \noalign{\medskip}{\textrm{e}^{\pm i
\phi_{k\ell}}}-A_{{k\ell}}\end {array} \right],
\end{equation}
where
\begin{equation}
	\gamma^\pm\,=\, 2-2\,\big(a_{k\ell}\cos\phi_{k\ell} \pm b_{k\ell}\sin\phi_{k\ell}\big)\cos\theta.
\end{equation}
For $\theta=\pi/3$ and $(k,l)=(0,0)$,  $\gamma^\pm=0$. In this case, the normalized eigenvectors are $\ket{v_{0 0}^{\pm\phi}}=\ket{\pm}$.

An eigenbasis of $U$ can be obtained from the eigenbasis of $U_{k \ell}$. In fact, the eigenvalues of $U_{k \ell}$ for $0\le k, \ell<n$ are exactly the  eigenvalues of $U$. On the other hand, a vector in the two-dimensional space is mapped to the Hilbert space ${\mathcal{H}}^{N}$  after multiplying its first entry by $\ket{{\psi}_{k \ell}^{\textrm{0}}}$ and its second entry by $\ket{{\psi}_{k \ell}^{\textrm{1}}}$. If $\ket{v_{k \ell}^\phi}$ is an eigenvector of  $U_{k \ell}$ associated with eigenvalue  $\exp({i{\phi_{k \ell}}})$ then the corresponding eigenvector of $U$ is
\begin{equation} \label{psi_kl}
 \ket{\psi_{k \ell}^\phi}= \braket{0}{v_{k \ell}^\phi}\,\ket{\psi^{0}_{k\ell}} +     \braket{1}{v_{k \ell}^\phi}\,\ket{\psi^{1}_{k\ell}} , \\  
\end{equation}
where $\ket{\psi^{0}_{k\ell}}$ and $\ket{\psi^{1}_{k\ell}}$ are given by Eqs.~(\ref{psikl0}) and~(\ref{psikl1}).

  Summing up, from the characterization  of the spectrum of $U_{k \ell}$,  we obtain an orthonormal eigenbasis of $U$, which is $\ket{\psi_{k \ell}^{\pm\phi}}$ for $0\le k,\ell<n$, where 
\begin{equation} \label{psi_kl_phi}
 \ket{\psi_{k \ell}^{\pm\phi}}= \frac{B_{{k\ell}}}{\sqrt{\gamma^\pm}}\ket{\psi^{0}_{k\ell}} + \frac{{\textrm{e}^{\pm i
\phi_{k\ell}}}-A_{{k\ell}}}{\sqrt{\gamma^\pm}}\ket{\psi^{1}_{k\ell}},\\  
\end{equation}
and the corresponding eigenvalues, which are $\textrm{e}^{\pm i\phi_{k\ell}}$ where $\cos\phi_{k\ell}=a_{k\ell}\cos{\theta}$.

\begin{figure}[h!] 
\centering
\includegraphics[trim=80 20 20 5,clip,scale=0.1]{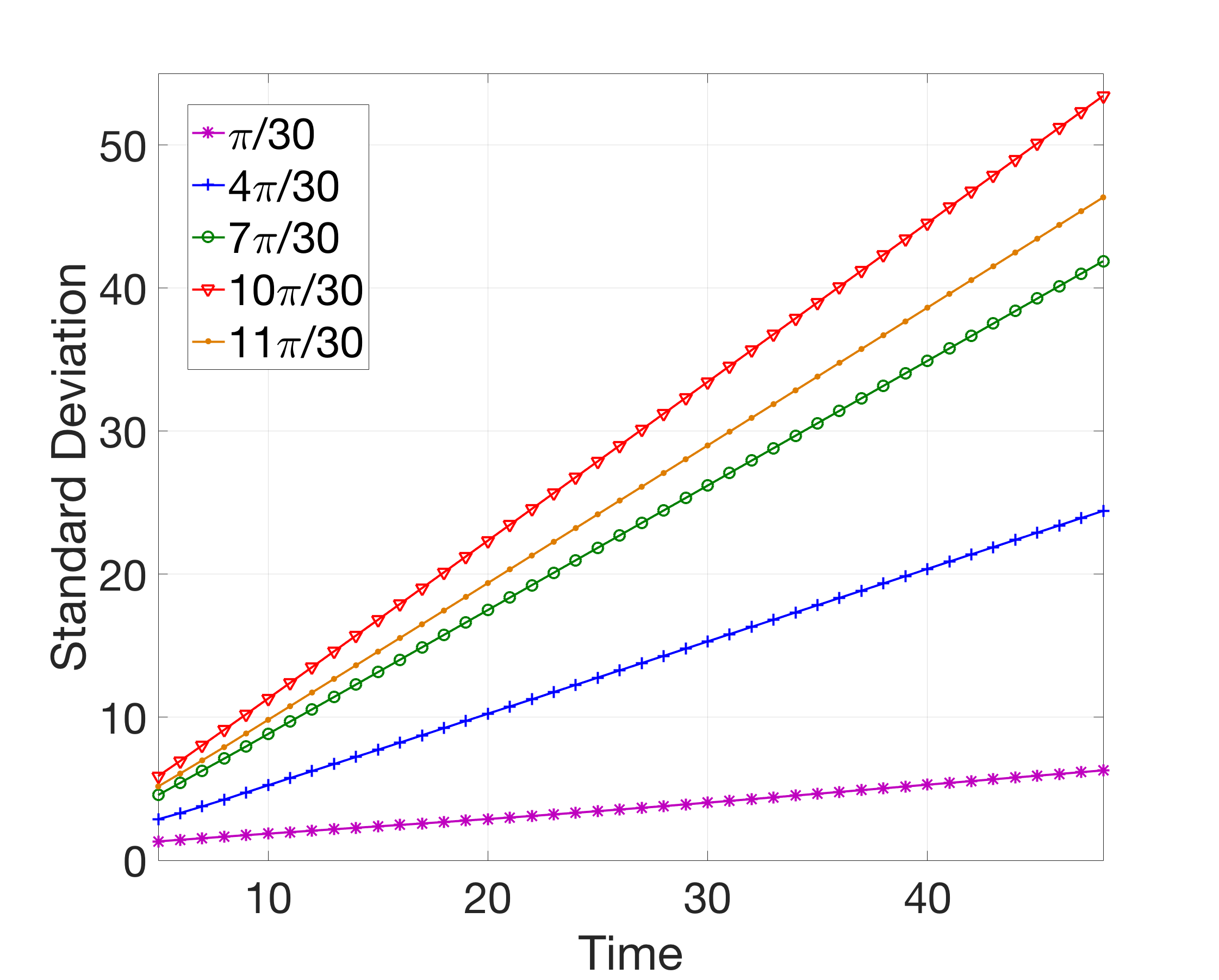}
\caption{Standard deviation $\sigma$ as a function of the number of steps $t$ for $\theta=\pi/30$, $4\pi/30$, $7\pi/30$, $\pi/3$, $11\pi/30$. The line with the largest slope has $\theta=\pi/3$.} 
\label{fig:slopes}
\end{figure}

\section{Standard Deviation}\label{sec:SD}

The standard deviation $\sigma(t)$ of the walker's position using the probability distribution $p_{x,y,i}(t)$ is an important quantity, which is expected to be proportional to $t$. In our numerical simulations, we implement the hexagonal lattice using regular hexagons considering each edge with one unit of length. We also take $n$ large enough to avoid the wave function to hit the lattice boundary.   Fig.~\ref{fig:slopes} shows  $\sigma$ as a function of $t$ for five values of $\theta$. Note that $\sigma(t)$ is in fact proportional to $t$ and the slope depends on $\theta$, for instance, the value $\theta=\pi/3$ results in the largest slope.

\begin{figure}[h!] 
\centering
\includegraphics[trim=80 20 20 5,clip,scale=0.1]{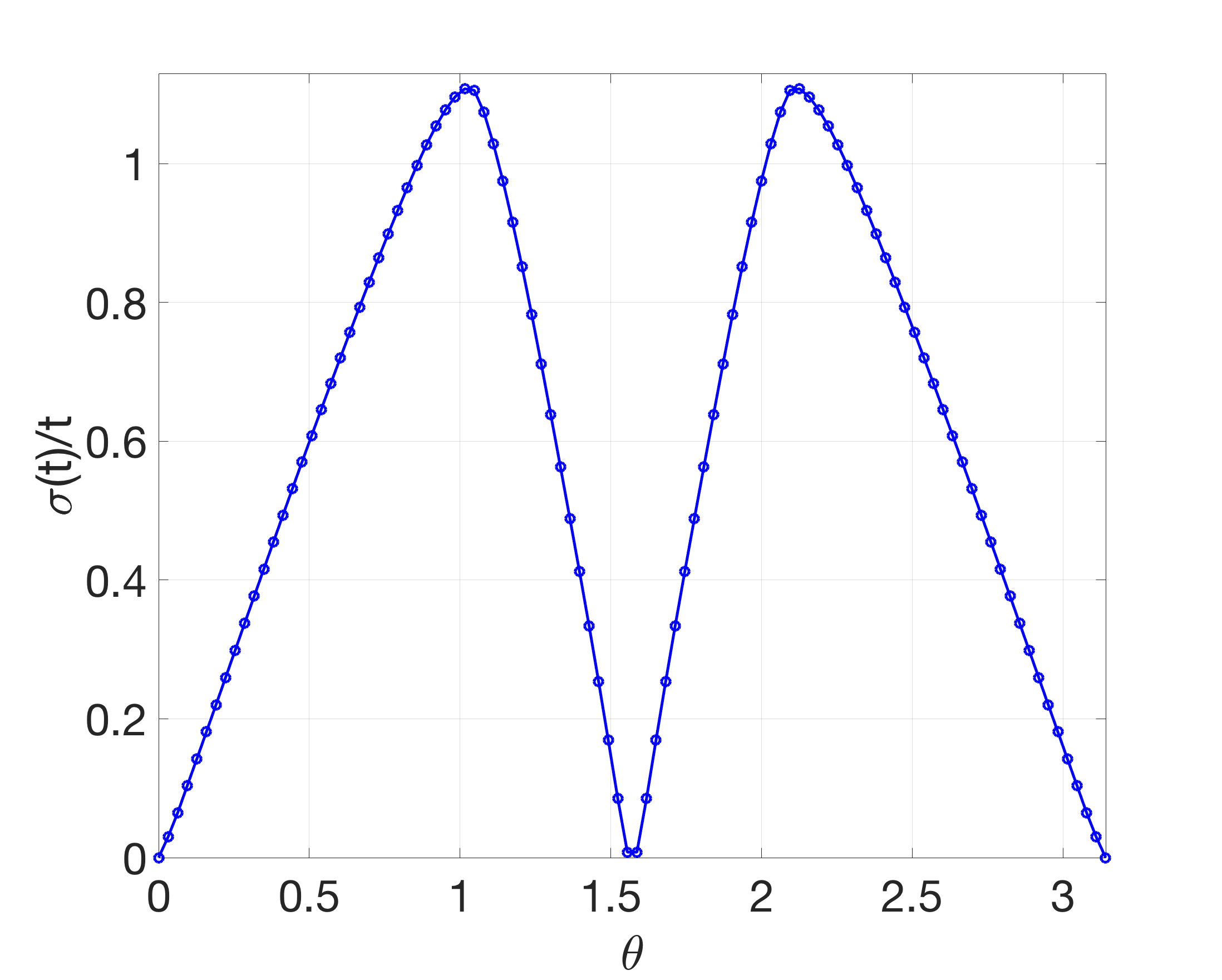}
\caption{$\sigma(t)/t$ as a function of $\theta$ for a large value of $t$.}
\label{fig:desvio}
\end{figure}

In order to obtain the optimal value of $\theta$, Fig.~\ref{fig:desvio} depicts $\sigma(t)/t$ as a function of $\theta$ for a sufficiently large value of $t$. The figure shows that there are three values of $\theta$ that result in a trivial dynamics, which are $\theta=0$, $\pi/2$, and $\pi$, and there are two optimal values, which are $\theta=\pi/3$ and $2\pi/3$ that have the maximal spreading rate. Those values coincide with the best values of $\theta$ for the spatial search algorithm as shown in the next section.

\section{No Localization}\label{sec:localization}

We say that a quantum walk exhibits localization at node $(x,y,j)$ if
\begin{equation}\label{eq:locdef_2} 
\limsup_{t\to\infty}\left|\braket{x,y,j}{\psi(t)}\right|^2 > 0,
\end{equation}
where $\ket{\psi(t)}=U^t\ket{\psi(0)}$ is the quantum walk state at time $t$.
Using the staggered Fourier basis, we have
\begin{equation}
 U^t = \sum_{k,\ell=0}^{n-1} \textrm{e}^{it\phi_{k\ell}}\ket{\psi_{k \ell}^{+\phi}}\bra{\psi_{k \ell}^{+\phi}} +\textrm{e}^{-it\phi_{k\ell}}\ket{\psi_{k \ell}^{-\phi}}\bra{\psi_{k \ell}^{-\phi}}.
\end{equation}
Using the results of Sec.~\ref{sec:FB}, we obtain
\begin{align}\label{eq:amplit}
\braket{x,y,j}{\psi(t)} = \frac{1}{n}\sum_{k,\ell=0}^{n-1} \left(h_{k\ell}^{+j}\textrm{e}^{it\phi_{k\ell}} + h_{k\ell}^{-j}\textrm{e}^{-it\phi_{k\ell}}  \right),
\end{align}
where 
\begin{equation}
  h_{k\ell}^{\pm 0}=   \frac{B_{k\ell}\omega^{kx+\ell y}}{\sqrt{\gamma^\pm}} \braket{\psi_{k \ell}^{\pm\phi}}{\psi(0)},
\end{equation}
and
\begin{equation}
  h_{k\ell}^{\pm 1}=   \frac{(\textrm{e}^{\pm i\phi_{k\ell}}-A_{k\ell})\omega^{kx+\ell y}}{\sqrt{\gamma^\pm}} \braket{\psi_{k \ell}^{\pm\phi}}{\psi(0)}.
\end{equation}
We suppose that $\ket{\psi(0)}$ is a superposition of a finite number of nodes and does not depend on $n$.

Note that usually there is localization at every vertex when $n$ is finite. It is interesting to analyze the localization of quantum walks on infinite lattices. We take the limit when $n$ approaches $+\infty$ and we convert the sums of Eq.~(\ref{eq:amplit}) into integrals as
$$\frac{1}{n}\sum_{k,\ell=0}^{n-1} \longrightarrow \frac{1}{(2\pi)^2}\int_{-\pi}^{\pi}\int_{-\pi}^{\pi},$$
and $\tilde{k} \rightarrow k$, $\tilde{\ell}\rightarrow \ell$.

The two-dimensional version of the stationary-phase method can be expressed by the following lemma~\cite{PW10}. 
\begin{lemma}
Let $f$ and $\phi$ from $[-\pi,\pi]^2$ to  $\mathbb{Cpr}$ be analytic functions. 
Assume that the set of the critical points $\Gamma$ of $\phi$ is finite, and the Hessian matrix $H=[\,\partial^2 f/\partial k \partial \ell]$ at each point in $\Gamma$ is non-singular. 
Then 
\begin{align}\label{eq:locdef}
\int_{-\pi}^{\pi}\int_{-\pi}^{\pi} f(k,l) \emph{e}^{\pm it\phi_{k\ell}}  {dk\,d\ell} \sim \hspace{2.5cm} \nonumber\\
\sqrt{\frac{1}{\pi t}} \sum_{(k_0,\ell_0)\in \Gamma} \frac{f(k_0,\ell_0)\emph{e}^{\pm it\phi_{k_0\ell_0}} }{\sqrt{\det H(k_0,\ell_0)}}.
\end{align}
\end{lemma}

To apply the above lemma, we take $f(k,l)=h_{k\ell}^{\pm j}$ and we have to check which are the critical points of $\phi(k,\ell)$. We have 
	\begin{align}
        \frac{\partial\phi_{k\ell}}{\partial k} &= \frac{\cos\theta\sin^2\theta\, \big(\sin(k-\ell)+\sin k\big)}{\sin \phi_{k\ell}}, \\
        \frac{\partial\phi_{k\ell}}{\partial \ell} &=  \frac{\cos\theta\sin^2\theta\, \big( \sin(\ell-k)+\sin \ell\big)}{\sin \phi_{k\ell}}.
        \end{align}
Then, there are 8 critical points in $\Gamma$ satisfying $\{(k,\ell)\in [-\pi,\pi] \;:\; \sin k+\sin (k-\ell)=\sin \ell+\sin(\ell-k)=0\}$.
The determinant of the Hessian matrix is 
\begin{align}
	 \det H(k,\ell) =\frac{\cos \theta\sin^2\theta }{\sin^2 \phi_{k,\ell}}\times\hspace{1.5cm} \nonumber\\
                	\bigg|  \begin{matrix} \cos k+\cos(k-\ell) & -\cos \phi_{k\ell}-\cos(k-\ell) \\ -\cos \phi_{k\ell}-\cos(k-\ell) & \cos \ell+\cos(k-\ell) \end{matrix}  \bigg|.
\end{align}	
It is straightforward to check that the Hessian matrix is is not degenerate at any critical point. Then, no vertex admits localization since the decay rate of right-hand side of (\ref{eq:locdef}) for every vertex is $O(1/t)$.

\section{Search}\label{sec:search}

The quantum search algorithm on hexagonal lattices is driven by a modified evolution operator
\begin{equation}
	{\mathcal{U}_0}\,=\,U\,R_0
\end{equation}
where $U$ is defined by Eq.~(\ref{eq:U}) with $\theta=\pi/3$ and $R_0$ is the unitary operator that inverts the sign of the marked vertices leaving the other vertices unchanged. We address the case with one marked vertex and without loss of generality we assume that the marked vertex is $(0,0,0)$. Then, $R_0=I-2\ket{0,0,0}\bra{0,0,0}$.

The initial condition is the uniform superposition of all vertices in order to avoid any bias towards the location of the marked vertex, that is
\begin{equation}
	\ket{\psi_0}\,=\,\frac{1}{\sqrt{2}\,n}\sum_{x,y=0}^{n-1} \big(\ket{x,y,0}+\ket{x,y,1}\big).
\end{equation}
The searching algorithm consists of applying ${\mathcal{U}_0}$ over and over until the probability of finding the marked vertex reaches its maximum, that is, the final state is $\ket{\psi_t}={\mathcal{U}_0}^t\ket{\psi_0}$, where $t$ is the optimal running time. Usually, the probability of finding the marked vertex after a measurement in the computational basis decreases when the system size increases. There are many strategies to boost the success probability. 

Our first task is to determine the optimal value of $\theta$. In order to simplify the calculations, we employ the spectrum of $(-U)$ instead of $U$. The eigenvalues of $(-U)$ are
$\textrm{e}^{\pm i(\pi+\phi_{k\ell})}$ where $\cos\phi_{k\ell}=a_{k\ell}\cos{\theta}$. The asymptotic expansion (for large $n$) of $\cos\phi_{k\ell}$ is
\begin{align}
	\left( 4\,  \cos^{2} \theta -3 \right) 
\cos  \theta  + & {\frac {4\,{\pi }^{2} \left( {k}^{2}-kl+{l
}^{2} \right)  \sin^{2} \theta \cos \theta }{{n}^{2}}} \nonumber  \\
+ & \, O \left( \frac{1}{n^4} \right). 
\end{align}
Let $\phi_\textrm{min}$ be the argument of the eigenvalue of $(-U)$ with the smallest positive argument, which is characterized by $(k,l)=(0,1)$, for instance. The above expansion shows that there is only one value of $\theta$ such that $\phi_\textrm{min}$ tends to 0 (we address only the range $0\le\theta\le\pi/2$). This special value is $\theta=\pi/3$. This means that for $\theta\neq\pi/3$ we have $\phi_\textrm{min}=\Theta(1)$ and the quantum search algorithm in the hexagonal lattice will be as slow as a random-walk-based search, which is $\Omega(N\ln N)$. For  $\theta=\pi/3$, we obtain $\phi_\textrm{min}=\sqrt{3}\pi/n+O(1/n^2)$.

Let $\exp(i \lambda)$ be the eigenvalue of ${\cal U}_0$ with the smallest positive argument $\lambda$ and let $\ket{\lambda}$ be its associated eigenvector, that is, ${\cal U}_0\ket{\lambda}=\exp(i \lambda)\ket{\lambda}$. We now describe a method to calculate $\lambda$ using the spectrum of $(-U)$. Recall that $(-U)$ is an evolution operator with no marked vertex.
Using the completeness relation and the fact that $\ket{\psi_{k \ell}^{\pm\phi}}$ is an orthonormal basis of the Hilbert space, we have
\begin{equation}\label{bra00lambda}
 \braket{0,0,0}{\lambda}\,=\,\sum_{k,\ell,\pm\phi} \braket{0,0,0}{\psi_{k \ell}^{\pm\phi}}\braket{\psi_{k \ell}^{\pm\phi}}{\lambda},
\end{equation}
where the sum runs over all values of $(k,\ell)$ for both $\pm\phi$. On the other hand, from the expression $\bracket{\psi_{k \ell}^{\phi}}{{\cal U}_0}{\lambda}=\bracket{\psi_{k \ell}^{\phi}}{U R_0}{\lambda}$, we obtain
\begin{equation}\label{psi_lambda}
 \braket{\psi_{k \ell}^{\pm\phi}}{\lambda}\,=\,\frac{2\braket{0,0,0}{\lambda}\braket{\psi_{k \ell}^{\pm\phi}}{0,0,0}}{1-\textrm{e}^{i(\pi+\lambda\pm\phi_{k \ell})}}.
\end{equation}
Using the above equation in (\ref{bra00lambda}) and $\braket{0,0,0}{\psi_{k \ell}^{\phi}}=\braket{0}{v_{k \ell}^{\phi}}/n$, we obtain
\begin{equation}
 \sum_{k,\ell,\pm\phi} \frac{\left|\braket{0}{v_{k \ell}^{\pm\phi}}\right|^2}{1-\textrm{e}^{i(\pi+\lambda\pm\phi_{k \ell})}} =\frac{n^2}{2},
\end{equation}
which is valid if $\lambda\pm\phi_{k \ell}$ is not an odd multiple of $\pi$ for all $k,\ell$. Using that $2/(1-\textrm{e}^{ix})=1+i\sin x/(1-\cos x)$, the imaginary part of the above equation is
\begin{equation}\label{exactsum_kl}
 \sum_{k,\ell,\pm\phi} \left|\braket{0}{v_{k \ell}^{\pm\phi}}\right|^2\frac{\sin({\lambda\pm\phi_{k \ell}})}{1+\cos(\lambda\pm\phi_{k \ell})} =0.
\end{equation} 
Separating the term $(k,\ell)=(0,0)$ from the above sum, we obtain
\begin{eqnarray}\label{sum3terms}
 \sum_{\substack{k,\ell=0  \\ (k,\ell)\neq(0,0) }}^{n-1} &&\left( \frac{\left|\braket{0}{v_{k \ell}^{+\phi}}\right|^2\sin({\lambda+\phi_{k \ell}})}{1+\cos(\lambda+\phi_{k \ell})}
+\right.\nonumber\\
&&\left.  \frac{\left|\braket{0}{v_{k \ell}^{-\phi}}\right|^2\sin({\lambda-\phi_{k \ell}})}{1+\cos(\lambda-\phi_{k \ell})}\right)
= \cot\frac{\lambda}{2} .
\end{eqnarray}
This equation can be used to calculate $\lambda$ numerically by choosing the positive solution closest to zero. To proceed analytically, we suppose that $\lambda\ll \phi_\textrm{min}$ for $n\gg 1$. We check the validity of this assumption in the next paragraph. By now we remark that $\left|\braket{0}{v_{k l}^{\phi}}\right|^2+\left|\braket{0}{v_{k l}^{-\phi}}\right|^2=1$.

Expanding Eq.~(\ref{sum3terms}) up to order $O(\lambda^2)$ we obtain
\begin{equation}\label{lambda_main}
\lambda\,=\,\pm\frac{1}{nC},
\end{equation}
where
\begin{eqnarray}\label{B2}
 C^2 &=& \frac{1}{n^2}\sum_{\substack{k,\ell=0  \\(k,\ell)\neq(0,0) }}^{n-1} 
\frac{1}{2+a_{k \ell}}+O(1),
\end{eqnarray}
and 
\begin{eqnarray}\label{aakl}
 a_{k\ell}=
\frac{3}{4}\left(\frac{1}{3}-\cos{\tilde{k}}-\cos{\tilde{\ell}}-\cos{\tilde{k}-\tilde{\ell}}\right),
\end{eqnarray}
when $\theta=\pi/3$. Eq.~(\ref{lambda_main}) shows that both $\exp(\pm i \lambda)$ are eigenvalues of ${\cal U}_0$. In the Appendix, we show that $C=\Theta(\sqrt{\ln n})$. Therefore, $1/\lambda=\Theta(\sqrt{N\ln N})$. Since $\phi_\textrm{min}=\Theta(1/\sqrt{N})$, this confirms that $\lambda\ll \phi_\textrm{min}$ for $n\gg 1$ is a consistent limit.

 Using~(\ref{psi_lambda}) in the normalization condition $\sum_{k l}  \left|\braket{\psi_{k l}^{\phi}}{\lambda}\right|^2=1$ and $\left|\braket{\psi_{k l}^{\phi}}{000}\right|^2=\left|\braket{0}{v_{k l}^{\phi}}\right|^2/n^2$, we obtain
\begin{align}
	\frac{1}{\left|\braket{000}{\lambda}\right|^2} = \frac{2}{n^2} \sum_{k l}  \frac{\left|\braket{000}{v_{k l}^{\phi}}\right|^2}{1+\cos\left(\lambda+\phi_{k l}\right)}+ \nonumber \\
	\frac{\left|\braket{000}{v_{k l}^{-\phi}}\right|^2}{1+\cos\left(\lambda-\phi_{k l}\right)}.
\end{align}
Separating the term with $(k,l)=(0,0)$, using that $\lambda\ll\phi_\textrm{min}$ for $n\gg 1$,  and keeping the dominant terms, we obtain
\begin{align}\label{ket00lambda_v2}
	\frac{1}{\left|\braket{000}{\lambda}\right|^2}=\frac{4}{n^2\lambda^2}+ \frac{4}{n^2}\sum_{\substack{k,l=0  \\(k,l)\neq(0,0)}}^{n-1}  \frac{1}{2+a_{k l} } \nonumber \\
	+   \,O\left(1\right).
\end{align}
Using Eqs.~(\ref{lambda_main}) and~(\ref{B2}), we obtain
\begin{equation}\label{ket00lambda}
	\frac{1}{\left|\braket{000}{\lambda}\right|^2} =\frac{8}{n^2\lambda^2}+O\left(1\right).
\end{equation}
Without loss of generality, we assume that $\braket{0,0,0}{\lambda}$ is positive and real. In fact, if $\braket{0,0,0}{\lambda}=a\,\textrm{e}^{ib}$, where $a$ and $b$ are real and $a$ is positive, we redefine $\ket{\lambda}$ as  $\textrm{e}^{-ib}\ket{\lambda}$. After this redefinition, $\braket{0,0,0}{\lambda}$ is a positive and real, given by $n\lambda/2\sqrt{2}+O(1)$. The same applies to $\braket{0,0,0}{\lambda^-}$, and we also obtain $\braket{0,0,0}{\lambda^-}=n\lambda/2\sqrt{2}+O(1)$.

Decomposing $\ket{\psi_0}$ in the eigenbasis of ${\cal U}_0$, we obtain
\begin{equation}\label{psi_0_lambda}
	\ket{\psi_0}= \braket{\lambda}{\psi_0}\,\ket{\lambda}+ \braket{\lambda^-}{\psi_0}\,\ket{\lambda^-}+\ket{\psi_0^\perp},
\end{equation}
where $\ket{\psi_0^\perp}$ is the component of $\ket{\psi_0}$ orthogonal to the plane spanned by $\ket{\lambda}$ and $\ket{\lambda^-}$, where $\ket{\lambda^-}$ is the eigenvector of ${\cal U}_0$ associated with $\exp(-\lambda i)$. Using Eq.~(\ref{psi_kl}) and $\ket{v_{0 0}^{\phi}}=\ket{+}$, we verify that $\ket{\psi_0}=\ket{\psi^{\phi}_{00}}$. Using Eq.~(\ref{psi_lambda}) with $(k,l)=(0,0)$, we obtain
\begin{equation}\label{lambda_psi}
	\braket{\lambda}{\psi_0}= \frac{\lambda-2\,i}{4} +O\left(\lambda^2\right).
\end{equation} 
Using Eq.~(\ref{psi_lambda}) with $(k,l)=(0,0)$ again, but this time replacing $\ket{\lambda}$ by  $\ket{\lambda^-}$, we obtain that $\braket{\lambda^-}{\psi_0}=\left(\braket{\lambda}{\psi_0}\right)^*$. It is straightforward to check that
\begin{equation}
\left|\braket{\lambda}{\psi_0}\right|^2+\left|\braket{\lambda^-}{\psi_0}\right|^2=\frac{1}{2}+O\left(\lambda^2\right).
\end{equation}
Then, we can ignore the term $\ket{\psi_0^\perp}$ in Eq.~(\ref{psi_0_lambda}).

Using (\ref{psi_0_lambda}) and (\ref{lambda_psi}), we obtain
\begin{eqnarray}
	{({\cal U}_0)^t}\ket{\psi_0} &=&  \left( \frac{(\lambda-2\,i)\,\textrm{e}^{i\lambda t}}{4}
		+O\left(\lambda^2\right)\right)\ket{\lambda}+\nonumber \\
	&& \left( \frac{(\lambda+2\,i)\,\textrm{e}^{-i\lambda t}}{4} +O\left(\lambda^2\right)\right)\ket{\lambda^-}.
\end{eqnarray}
Using (\ref{ket00lambda}), we obtain
\begin{equation}\label{ket00Upsi0}
	\left|\bracket{000}{({\cal U}_0)^t}{\psi_0}\right|^2=
	\frac{n^2\lambda^2}{8}\, \sin^2(\lambda\,t)+O\left(\lambda^2\right).
\end{equation}

The running time is the first value of $t$ that maximizes the right-hand side of Eq.~(\ref{ket00Upsi0}), which is
\begin{equation}
	t\,=\, \frac{\pi}{2\lambda},
\end{equation}
ignoring terms $O(\lambda^2)$. The success probability is
\begin{equation}
	P = \frac{n^2\lambda^2}{8}+O\left(\lambda^2\right).
\end{equation}
Since $1/\lambda=\Theta(\sqrt{N\ln N})$, the running time is $t=\Theta(\sqrt{N\ln N})$ and the success probability is $P=\Theta(1/\ln N)$.

Tulsi~\cite{Tul12} described a modification of spatial search algorithms, which can be used to boost the success probability without increasing the number of steps. By employing this modification, the success probability is $O(1)$ and the number of steps is still $O(\sqrt{N\ln N})$.

\section{Conclusions}\label{sec:conc}

We have analyzed the dynamics of the staggered quantum walk with Hamiltonians on the hexagonal lattice with the focus on the position standard deviation, localization, and searching. The evolution operator has the parameter $\theta$ that can be tuned in order to produce the maximum spread rate of the wave function and the quickest spatial search algorithm. The best value of $\theta$ is $\pi/3$ for both cases. There are at least two interesting questions: (1)~The parameters that produce the highest spread rate does coincide with the parameters that produce the quickest search algorithm? (2)~In searching algorithms, is $\pi/k$ the optimal value of $\theta$, where $k$ is the number of tessellations?

We have also shown that the staggered quantum walk on the infinite hexagonal lattice does not admit localization. Note that this property cannot be used to conclude that this model is better for quantum transport or spatial searching because the analysis was performed using the non-modified evolution operator. It is an interesting research problem to check whether the modified evolution operator admits localization at some vertex for some initial condition.

\section*{Acknowledgments}
BR and RP acknowledge financial support from CNPq and CAPES. SB acknowledges financial support from CNPq through the “Ciência sem Fronteiras” program and thanks LNCC for its hospitality. E.S. acknowledges financial support from the Grant-in-Aid for Young Scientists (B) and of Scientific Research (B) Japan Society for the Promotion of Science (Grants 16K17637 and 16K03939).

\

\

\section*{Appendix}

Now we show that $C=\Theta(\sqrt{\ln n})$. Using Eqs.~(\ref{B2}), (\ref{aakl}), and trigonometric identities, we obtain
\begin{equation}\label{C2ineq}
\frac{1}{n^2}\sum_{\substack{k,\ell=0  \\(k,\ell)\neq(0,0) }}^{\frac{n}{2}-1} 
\frac{1}{2+a_{k \ell}}
\le 
C^2 
\le 
\frac{4}{n^2}\sum_{\substack{k,\ell=0  \\(k,\ell)\neq(0,0) }}^{\frac{n}{2}-1} 
\frac{1}{2+a_{k \ell}}
\end{equation}
Using that 
\begin{equation}
1-\frac{2\pi^2 k^2}{n^2} \le \cos\left(\frac{2\pi k}{n}\right)\le 1-\frac{8k^2}{n^2}
\end{equation}
for $0\le k<n/2$, we obtain
\begin{align}
\frac{1}{3\pi^2}\sum_{\substack{k,\ell=0  \\(k,\ell)\neq(0,0) }}^{\frac{n}{2}-1} 
\frac{1}{k^2+\ell^2-k\ell}
\le
C^2
\le \nonumber\\
\frac{1}{12}\sum_{\substack{k,\ell=0  \\(k,\ell)\neq(0,0) }}^{\frac{n}{2}-1} 
\frac{1}{k^2+\ell^2-k\ell}
\end{align}
Using that 
\begin{equation}
\frac{k^2+\ell^2}{2} \le k^2+\ell^2-k\ell \le k^2+\ell^2,
\end{equation}
we obtain
\begin{equation}
\frac{S(n)}{3\pi^2}
\le
C^2
\le \frac{S(n)}{6},
\end{equation}
where
\begin{equation}
S(n)=
\sum_{\substack{k,\ell=0  \\(k,\ell)\neq(0,0) }}^{\frac{n}{2}-1} 
\frac{1}{k^2+\ell^2}.
\end{equation}
The sum $S(n)$ has been calculated asymptotically in~Ref.~\cite{AKR05}, which proved that $S(n)$ is $\Theta({\ln n})$. This shows that $C=\Theta(\sqrt{\ln n})$.


\

\end{document}